\documentclass[12pt]{amsart}
\usepackage{amsmath, amssymb, amscd}
\usepackage{graphicx}
\usepackage{float}

\def\@cite#1{$^{#1}$} 
\def\newblock{\hskip .11em plus.33em minus.07em               
\sloppy\clubpenalty4000\widowpenalty4000                      
\sfcode`\.=1000\relax}

\begin{document}  


\title{Dynamics of Ebola epidemics in West Africa 2014}


\author{Robin J. Evans} \author{Musa Mammadov} 
\address{University of Melbourne, Parkville VIC 3010, Australia}

%
%

\begin{abstract}
This paper investigates the dynamics of Ebola virus transmission in West Africa during 2014. The reproduction numbers for the total period of epidemic and for different consequent time intervals are estimated based on a simple linear model. It contains one major parameter - the average time of infectiousness that defines the dynamics of epidemics.

Numerical implementations are carried out on data collected from three countries Guinea, Sierra Leone and Liberia as well as the total data collected worldwide. Predictions are provided by considering different scenarios involving the average times of infectiousness for the next few months and the end of the current epidemic is estimated according to each scenario.
\end{abstract}

\maketitle

\markboth{}{}


\noindent {\bf Funding Statement}\\
This work was funded by the Collaborative Research Network (CRN) of the Federation University of Australia anf the University of Melbourne.

\section{Introduction}
The outbreak of the 2014 Ebola virus epidemic in West Africa started in late 2013.There is currently no sign of the epidemic stopping, only evidence of things become worse. According to some estimates, by early December 2014 ``10,000 new cases of Ebola a week are possible" (\cite{Bosely}).

The epidemic does not seem to be under control and accurate predictions appear to be extremely difficult. The major reason for this might be due to unstable treatment conditions that provide different reproduction numbers at different periods. However, there are also other challenges related to the mathematical modeling of this epidemic. To address these challenges, several new models have been suggested that show quite different results, we note a few of them published recently \cite{althaus2014estimating,browne2014model,chowell2004basic,chowell2014transmission,nishiura2014early,
rivers2014modeling,periods2014ebola}.

In this article we introduce a new model to study the dynamics of epidemics by considering
the average time from onset to hospitalization as a time-dependent parameter. It is derived from the well studied $SIR$ (Susceptible-Infectious-Recovery) model, where the decrease in the number of susceptible population in compartment $S$ is the major force stopping epidemics. The susceptible population $S$ is often considered as a whole population.
A major drawback of this model, in terms of the current epidemic, is that the population infected constitutes a very small proportion of the total population, a very small decrease in $S$ has almost no effect on compartment $I.$

We discuss how this drawback could be tackled and introduce a new model that uses only compartment $I.$
This leads to a linear model having some similarities to those models based only on transmission rates from infectious population
at different generations (e.g. \cite{nishiura2014early}).
Our main goal is to fit data by estimating fewer but the most influential parameters without considering many other
issues like the infectiousness in hospitals and death ceremonies.
This in addition, allows us to have a more robust model with easily interpreted
parameters that can be used for more reliable predictions.
The main parameter in this model is the average infectiousness period $\tau_2$
(time from onset to hospitalization) that defines the dynamics of infectious population. This parameter
can also be considered as a control parameter in the development of
control models dealing with the spread of infection.

We calculate the basic reproduction numbers $\text{R}_0$ for each country (Guinea, Sierra Leone and
Liberia) as well as the total Ebola data worldwide. We
also provide predictions corresponding to different scenarios by considering different values for $\tau_2$ for future time periods.

\section{Methods}
We use the notation $I_a(t)$ for the number of "active" infectious population at time $t;$ it
represents the the total number of infectious population that are not yet hospitalized.
$C(t)$ and $D(t)$ are the cumulative number of infected cases and deaths, respectively.
The population density of a country is denoted by $\mathcal{D}.$ This is used in the definition of the
infection force of the disease with coefficient $\beta.$
Moreover, $\mu$  stands for the natural death rate of the population,
$\alpha$ for the death rate due to disease, and
$\tau_1$ for the average latent period (in days) that infected individuals become infectious
and $\tau_2$ for the average infectiousness period (in days).\\

The main equations of our model are as follows (see Appendix for details):
\begin{equation}\label{I_a_0}
     I_a(t+1) =  (1-\mu)^{\tau_1}  \sum_{i = 0}^{\tau_2-1} (1 - \mu)^i (1-\alpha \omega(i)) \, \beta \, \mathcal{D} \, I_a(t-\tau_1 - i);
\end{equation}
\begin{equation*}
     C(t+1) =  (1-\mu)^{\tau_1} \sum_{s = 1}^{t} \beta \, \mathcal{D} \, I_a(s-\tau_1);
\end{equation*}
\begin{equation*}
     D(t+1) =  (1-\mu)^{\tau_1} \sum_{s = 1}^{t}  \sum_{i = 0}^{n} (1 - \mu)^i\, \alpha\, \omega_p(i) \, \beta \, \mathcal{D} \, I_a(s-\tau_1 - i).
\end{equation*}

Here $\omega$ is a gamma (cumulative) distribution function (with p.d.f - $\omega_p$) for deaths due to disease \cite{periods2014ebola}; for the values of the parameters see Appendix.
We note that there are only three parameters that need to be estimated
to fit data for cumulative number of infected and death cases. These parameters are:
\begin{itemize}
  \item $\alpha$ - the death rate due to disease;
  \item $\beta$ - the coefficient of the force of infection;
  \item $\tau_2$ - the time to hospitalization.
\end{itemize}

Here $\alpha$ and $\beta$ are continuous variables, $\tau_2$ is a discrete variable with integer values (days).\\

{\bf Basic Reproduction Number  - $R_0$.~}
We calculate the basic reproduction number by considering the stationary states in (\ref{I_a_0}) as follows:
\begin{equation}\label{R_0}
    R_0 = \beta \, \mathcal{D} (1-\mu)^{\tau_1}  \sum_{i = 0}^{\tau_2-1} (1 - \mu)^i (1-\alpha \omega(i)),
\end{equation}
Since the natural death rate $\mu$ is close to zero; that is, $1-\mu \approx 1,$ from (\ref{R_0}) we have
\begin{equation*}\label{R_0_1}
    R_0 \approx \beta \, \mathcal{D} \, [\tau_2 - \alpha \sum_{i = 1}^{\tau_2-1} \omega(i)].
\end{equation*}
Moreover, since $\alpha \sum_{i = 0}^{\tau_2-1} \omega(i) < 1,$ the reproduction number $R_0 \approx \beta \, \mathcal{D} \tau_2.$
This means that the reproduction number depends almost linearly on $\tau_2$ - the time from onset to hospitalization.\\

{\bf The effective reproduction numbers - $R_k, k \ge 1.$~}
The effective reproduction numbers $R_k$ are considered on several consecutive time intervals
$\Delta_k = [t_k, t_{k+1}],$ $k = 1, 2, \cdots,$
with corresponding values of $\tau_2$. They are calculated by the same formula as $R_0.$

Here we make a reasonable assumption that the transmission rate $\beta \, \mathcal{D}$ describes the interaction of population (that is, in some sense, related to the local conditions and the life style) and should remain relatively stable for a particular country.
Then, the efforts in preventing the spread of infection are mainly observed in the change (decrease) in
the value of $\tau_2.$

Therefore, to calculate the effective reproduction numbers, we fit data and find the optimal values for $\alpha$ and $\beta,$ that are constant over the whole period, and
optimal values $\tau_2^k$ on each interval $\Delta_k.$ Then $R_k$ is calculated
by formula (\ref{R_0}) setting $\tau_2 = \tau_2^k.$\\

The sequence of optimal values $\tau_2^1,$ $\tau_2^2, \cdots,$
is considered as a method to describe the effectiveness of measures applied for preventing the spread of infection. This sequence very much defines the reproduction numbers on each consecutive time interval and therefore the dynamics of the infected population. It also allows us to consider future scenarios in terms of possible average times for hospitalization.

\section{Results and Discussion}

Data were retrieved from the WHO website for the cumulative numbers of clinical cases (confirmed, probable and suspected) collected till 11 November 2014. In all numerical experiments, the second half of the available data for each country is used for fitting the cumulative numbers of infected cases and deaths. The global optimization algorithm DSO in Global And Non-Smooth Optimization (GANSO) library \cite{Ganso, Mam-2005-Chapter} is applied for finding optimal values of parameters.

\begin{figure}[h!]
\caption{{\scriptsize The best fits for the cumulative numbers of infected cases and deaths in Guinea, Sierra Leone,
Liberia and worldwide by considering three parameters $\alpha$, $\beta$ and $\tau_2$ (for the values see Table \ref{Table_R_0}).
The lines represent the best fits, red and black circles represent the data.}} \label{Fig01}
\centering
\begin{tabular}{cc}
\includegraphics[scale=0.45]{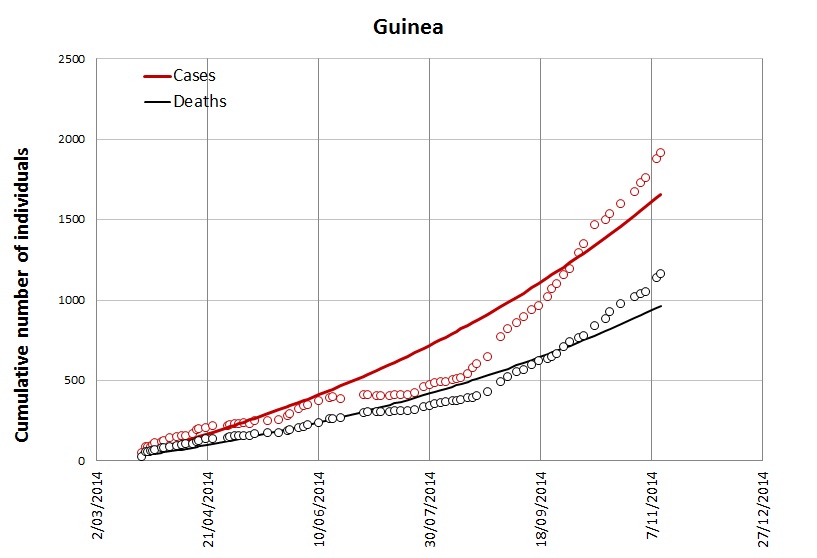} & \includegraphics[scale=0.45]{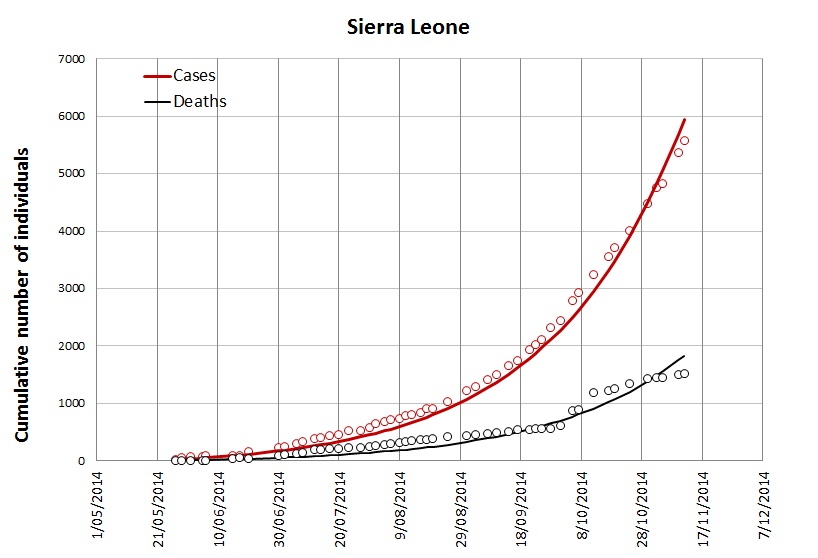}\\
\includegraphics[scale=0.45]{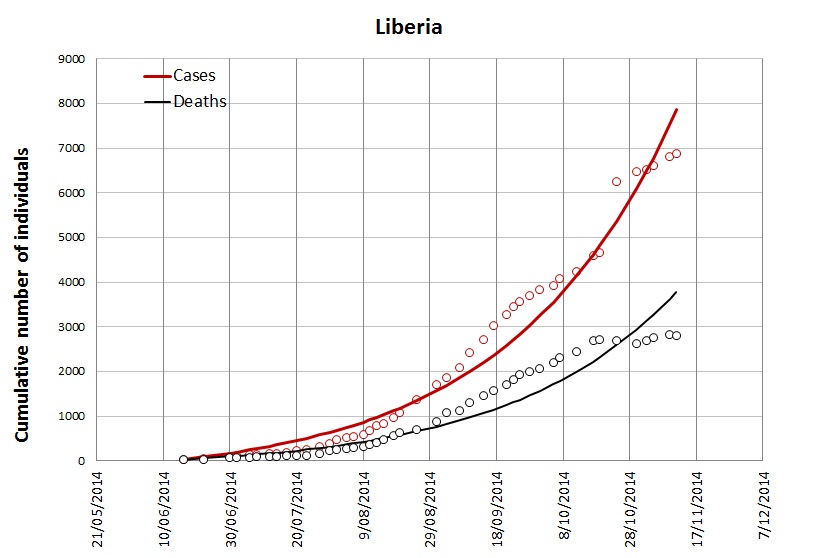} & \includegraphics[scale=0.45]{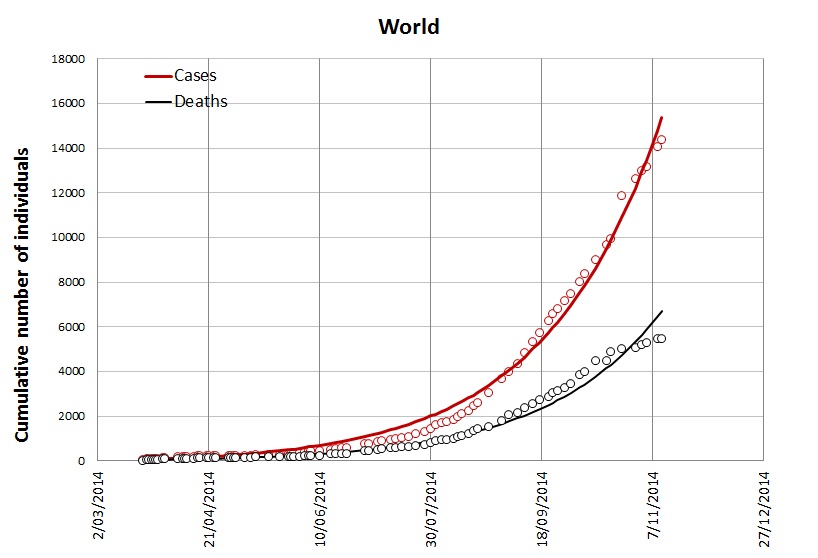}
\end{tabular}
\label{plots}
\end{figure}

First we consider the whole period of infection in each country and find the best fit in terms of three variables $\alpha$, $\beta$ and $\tau_2$ (Problem ($DF_1$) in Appendix). The results are presented in Table \ref{Table_R_0}. Although from Figure \ref{Fig01} it can be observed that the best fit for Guinea is not as good as for the other cases, these results provide some estimate for the reproduction number $R_0$ for a whole period of infection till 11-Nov-2014. In all cases (except Guinea), $R_0$ is between 1.20-1.23 and for Guinea - 1.06. We note that the dynamics of infected population is much more complicated which suggests that the reproduction number has been changing since the start of Ebola-2014 in almost all countries. This fact has been studied in \cite{periods2014ebola} (Firure S7) in terms of the instantaneous reproduction number over a 4-week sliding windows for each country (see also the next section for different values for $\tau_2$).

\begin{table}
\caption{{\small Results of best fits: optimal values for parameters $\alpha$, $\beta$ and $\tau_2.$
$R_0$ is the reproduction number.} }\label{Table_R_0}
\medskip
\centering
{\small 
\begin{tabular}{|c|ccc|c|}
  \hline
Country & $\alpha$ & $\beta$ & $\tau_2$ (days) & $R_{0}$\\
\hline
Guinea       & 0.603  &   0.00311  &  10 &     1.064 \\
Sierra Leone & 0.364  &   0.00511  &   3 &     1.227 \\
Liberia      & 0.555  &   0.01108  &   3 &     1.196 \\
World        & 0.501  &   0.00362  &   7 &     1.206 \\
 \hline
\end{tabular}
}
\end{table}

\subsection{The effective reproduction numbers}

According to (\ref{R_0}), the basic reproduction number is mainly determined by $\beta$ and $\tau_2.$
Since in our model parameter $\tau_2$ takes discrete values (days) it would be interesting to study
the change of this parameter over time while keeping $\beta$ the same for the whole period. This approach makes it possible to consider different scenarios for future developments regarding the change in this parameter and to provide corresponding predictions.

\begin{figure}[h!]
\caption{{\scriptsize The best fits for the cumulative numbers of infected cases and deaths in Guinea, Sierra Leone,
Liberia and worldwide by considering three parameters $\alpha$, $\beta$ and three subintervals
with different values $\tau_2^k,$ $k=1,2,3$ (for the values see Table \ref{Table_R_k}).
The lines represent the best fits, red and black circles represent the data.}} \label{Fig02}
\centering
\begin{tabular}{cc}
\includegraphics[scale=0.45]{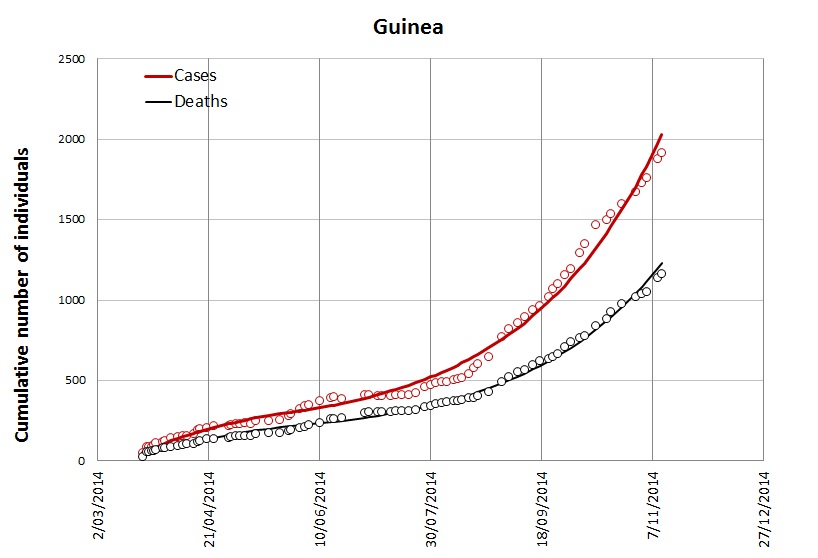} & \includegraphics[scale=0.45]{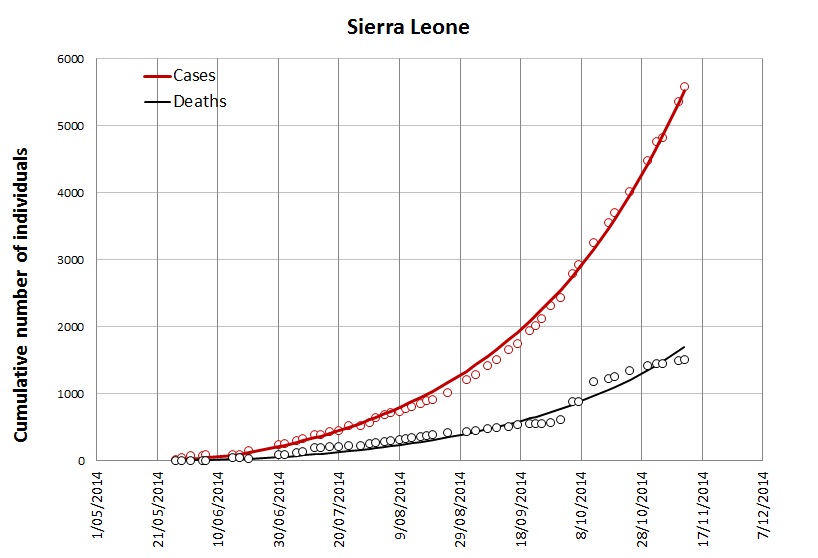}\\
\includegraphics[scale=0.45]{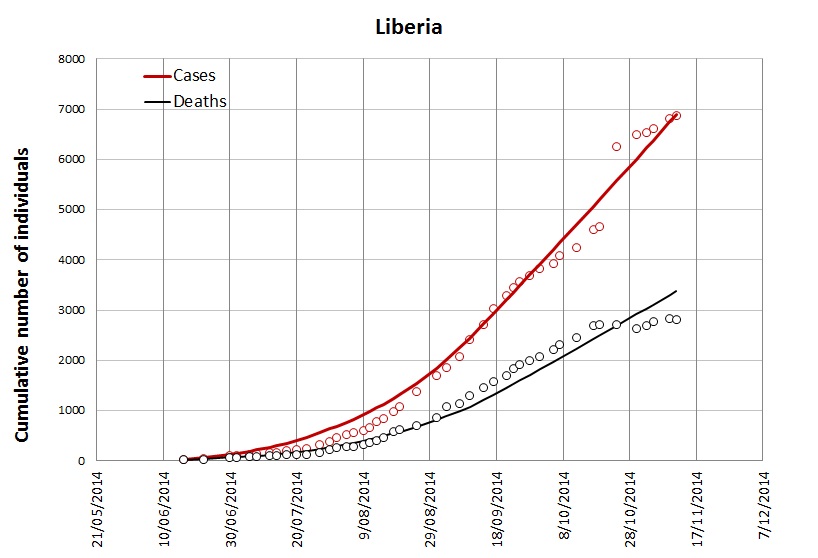} & \includegraphics[scale=0.45]{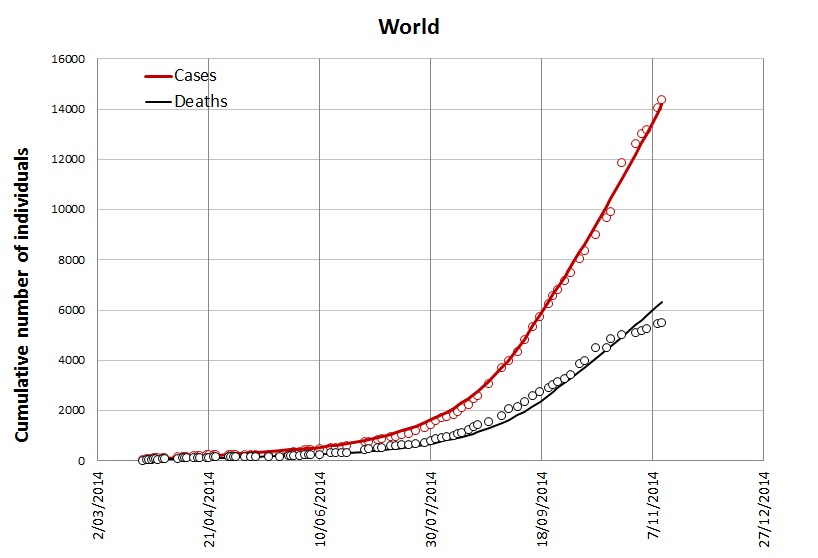}
\end{tabular}
\label{plots}
\end{figure}

We consider three consequent intervals $\Delta_k = [t_k, t_{k+1}]$ $(k = 1,2,3)$ for each country and find optimal values
$\alpha,$ $\beta$ and $\tau_2^k$ $(k = 1,2,3)$ (Problem ($DF_2$) in Appendix). The results are presented in Table \ref{Table_R_k}.
The last time point $t_4$ is 11-Nov-2014. The values of $t_1, t_2,t_3$ are as follows:
22-March, 23-May and 20-July for Guinea;
27-May, 20-June and 20-August for Sierra Leone;
16-June, 20-July and  07-Sept for Liberia; and
22-March, 23-May and  07-Sept for the total data (World).
Each interval $\Delta_k$ has its own reproduction number $R_k$ that defines the shape of the best fits
presented in Figure \ref{Fig02}.

In all cases the effective reproduction number is still grater than 1. In Liberia it shows a decrease from 1.48 to 1.01 and this can be seen in quite a noticeable decrease in the number of cumulative infected cases (Figure \ref{Fig02}).

\begin{table}
\caption{{\small Results of best fits: the (effective) reproduction numbers $R_k$ and average times to hospitalization $\tau_2^k$ (in days) for different intervals $\Delta_k,$ $k = 1,2,3.$ The optimal values for $\alpha$ and $\beta$ are also provided; they are constant for a whole period}}\label{Table_R_k}
\medskip
\centering
{\small 
\begin{tabular}{|c|cc|ccc|}
  \hline
Country & $\alpha$ & $\beta$ & $R_1$ ($\tau_2^1$)  & $R_2$ ($\tau_2^2$)& $R_3$ ($\tau_2^3$) \\
\hline
Guinea       & 0.661 & 0.00506 & 0.827 (4)  & 1.203 (6) & 1.203 (6) \\
Sierra Leone & 0.350 & 0.00368 & 1.736 (6)  & 1.176 (4) & 1.176 (4) \\
Liberia      & 0.529 & 0.00705 & 1.482 (6)  & 1.255 (5) & 1.012 (4) \\
World        & 0.489 & 0.00519 & 1.035 (4)  & 1.285 (5) & 1.035 (4) \\
 \hline
\end{tabular}
}
\end{table}

\subsection{Future scenarios}
We consider only the cumulative number of infected population worldwide.
From Table \ref{Table_R_k} it can be observed that the number $\tau_2$ has changed as 4, 5 and 4 from 22-March to 11-Nov.
We keep this initial best fit (the optimal values of parameters (World) are in Table \ref{Table_R_k})
and consider different scenarios for possible changes of this parameter in the future
while keeping the values of $\alpha$ and $\beta$ unchanged.

The future time intervals are designed as follows: the first interval $\Delta_1$ is 12-Nov-2014-31-Dec-2014, followed by each next month
$\Delta_2 - \Delta_4,$ and the last interval $\Delta_5$ starts from 1-Apr-2015.
The results are presented in Table \ref{Table_W_prediction}.
The reproduction numbers are 0.778 (for $\tau_2$ = 3), 1.035 (for $\tau_2$ = 4) and 1.284 (for $\tau_2$ = 5).

In the best scenario in Table \ref{Table_W_prediction} it is assumed that
the current trend stays stable ($\tau_2 = 4$) and a 25 percent decrease in the hospitalization time starts from 1-Jan-2015,
then the epidemic may continue till Apr-2015 with the total number of infected cases reaching 31,000.

The worst case considered in Table \ref{Table_W_prediction} assumes that during the next two months
(from 12-Nov-2014 to 31-Jan-2015)
the average time to hospitalization increases by 25 percent (that is, from 4 days to 5 days) and
then gradually decreases in Feb-Mar-2015 (from 5 days to 4 days), in Apr-2015 (from 4 days to 3 days)
and stays at this level afterwards. In this case,
the Ebola outbreak could be stopped by July-2015 with the total number of infected cases reaching 165,000.

\begin{table}
\caption{{\small The cumulative number of infected population according to different scenarios corresponding different values $\tau_2^k$ for time intervals $\Delta_k.$ The starting values of parameters ($\alpha$, $\beta$ and $\tau_2^k,$ $k=1,2,3$) are in Table \ref{Table_R_k} (World). The first time interval ($\Delta_1$) is 12/Nov/2014-31/Dec/2014, followed by each next month and the last interval ($\Delta_5$) starts from 1/Apr/2015. The last column presents the date for the end of Ebola epidemic (see also Figure \ref{Fig03}) with corresponding number of cumulative infected population $C_{max}$ (-/$\infty$ means no stabilization). The version  $\tau_2^k = 4$ for all $k$ means the current trend remains unchanged.
The reproduction number for $\tau_2 = 3$ is $R = 0.778;$ it is less than 1 which leads to stabilization.
For corresponding reproduction numbers for $\tau_2 = 4$ and 5 see Table \ref{Table_R_k} (World).} }\label{Table_W_prediction}
\medskip
\centering
{\small 
\begin{tabular}{|c|c|c|c|c|c|c|}
  \hline
$\tau_2^1$ & $\tau_2^2$ & $\tau_2^3$ & $\tau_2^4$ & $\tau_2^5$ &  End/$C_{max}$\\
\hline
4 & 4 & 4 & 4 & 4  &  -/$\infty$\\
4 & 3 & 3 & 3 & 3  &  Apr-2015/31,000\\
4 & 4 & 3 & 3 & 3  &  Apr-2015/39,000\\
4 & 4 & 4 & 3 & 3  &  May-2015/47,000\\
4 & 4 & 4 & 4 & 3  &  May-2015/57,000\\
4 & 5 & 4 & 3 & 3  &  Jun-2015/69,000\\
4 & 5 & 4 & 4 & 3  &  Jun-2015/90,000\\
4 & 5 & 5 & 4 & 3  &  Jun-2015/135,000\\
5 & 4 & 4 & 3 & 3  &  Jun-2015/102,000\\
5 & 5 & 4 & 3 & 3  &  Jul-2015/120,000\\
5 & 5 & 4 & 4 & 3  &  Jul-2015/166,000\\
\hline
\end{tabular}
}
\end{table}

\begin{figure}[h!]
\caption{{\scriptsize The cumulative number of infected population according
to different scenarios corresponding different values $\tau_2^k$ for time intervals
$\Delta_k.$ The starting values of parameters ($\alpha$, $\beta$ and $\tau_2^k,$ $k=1,2,3$)
are in Table \ref{Table_R_k} (World). The first time interval ($\Delta_1$) is 12/Nov/2014-31/Dec/2014,
followed by each next month and the last interval ($\Delta_5$) starts from 1/Apr/2015.
The reproduction number for $\tau_2 = 3$ is $R = 0.778;$ it is less than 1 which leads to stabilization.
For corresponding reproduction numbers for $\tau_2 = 4$ and 5 see Table \ref{Table_R_k} (World).}} \label{Fig03}
\centering
\begin{tabular}{cc}
\includegraphics[scale=0.42]{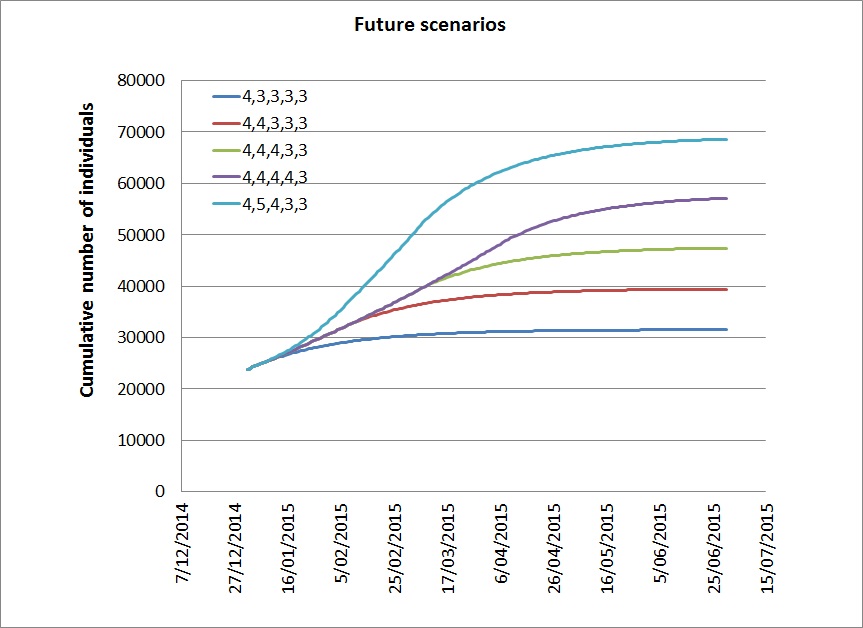} & \includegraphics[scale=0.42]{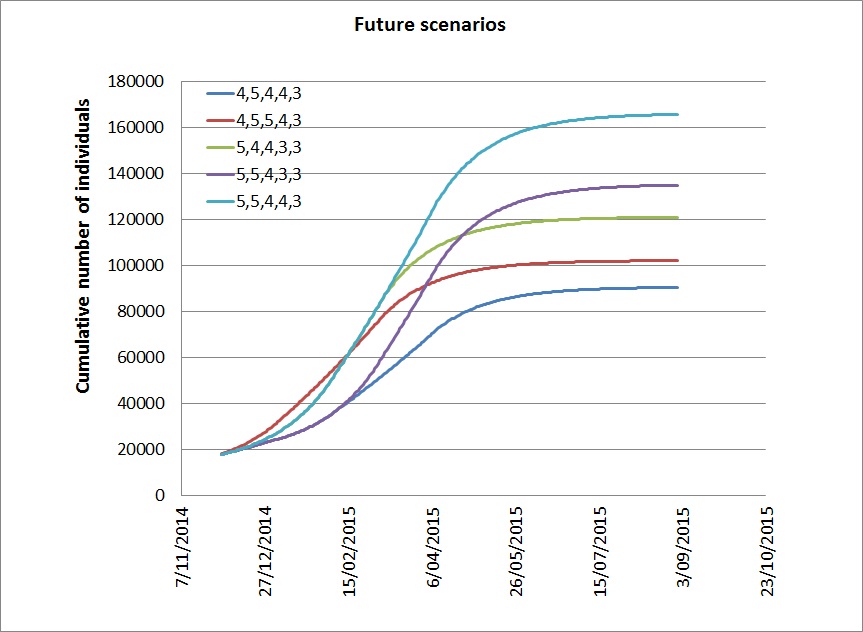}
\end{tabular}
\label{plots}
\end{figure}



\bibliographystyle{plain}


\newpage

{\Large {\bf APPENDIX}}\\

\section{Model}
 The idea behind the model introduced in this paper is related to the  $SIR$  model with time delay. Since we are going to implement it on available daily data, a discrete version of this model is considered with the step-size one day. Moreover, since the "recovered" population is not our focus, we will only consider equations related to susceptible ($S$) and infectious ($I$) individuals. The most commonly used $SIR$ model\footnote{Continuous time version of this model is
\begin{equation*}
    \frac{dS(t)}{dt} = \lambda - \mu S(t) - F(S(t),I(t)), ~~~
    \frac{dI(t)}{dt} = e^{-\mu \tau_1} F(S(t-\tau_1),I(t-\tau_1)) - (\mu + \alpha + \gamma) I(t)
\end{equation*}
}
in the literature is provided below (see, for example, \cite{li2014sir}):
\begin{equation}\label{S}
    S(t+1) = \lambda + (1-\mu) S(t) - F(S(t),I(t));
\end{equation}
\begin{equation}\label{I_A}
     I(t+1) = (1-\mu)^{\tau_1} F(S(t-\tau_1),I(t-\tau_1)) + (1 - \mu - \alpha - \gamma) I(t).
\end{equation}
Here
$\lambda$ is the recruitment of the population;
$\mu$  is the natural death rate of the population;
$\alpha$ is the death rate due to disease;
$\gamma$ is the recovery rate; and
$\tau_1$ is the latent period that infected individual becomes infectious.

The fraction $(1-\mu)^{\tau_1}$ represents the survival rate of population over the period of $[0,\tau_1]$ (in continuous-time case it is equivalent to $e^{-\mu \tau_1}$). Below we examine this model in detail and develop an improved model.\\

{\bf Susceptible individuals.~} Equation (\ref{S}) describes the dynamics of susceptible individuals $S(t).$
This equation "keeps" the number of infectious individuals $I(t)$ bounded.
For example, when the basic reproduction number is greater than 1, there exists \cite{li2014sir}
an endemic equilibrium   $(S^\star,I^\star)$   and  $S(t) \to S^\star,$  $I(t) \to I^\star$ as $t \to \infty.$
In the case when the birth rate is zero ($\lambda=0$) the relation  $I(t) \to 0$  suggests that $S(t) \to 0.$

Thus according to this model the epidemic ends because the number of
susceptible individuals $S(t)$  decreases over time and the effective reproduction number (as a function of time)
becomes less than 1 at some stage;
that is, the number of newly infected population $F(S(t),I(t))$ decreases thanks to the "enough" decrease in
the number of susceptible individuals (while $I(t)$ still increases).
This might be applicable to epidemics in early 1900s but it is definitely not applicable to recent ones.

This issue significantly restricts the application of the $SIR$ model for the study of the current Ebola virus epidemic.
Below we consider 3 possibilities to overcome this difficulty.\\

{\bf 1.} The simplest way would be to
use a "relatively small" number $S(0)$ for a possible number of susceptible individuals
that may become infected. This approach has been implemented
in \cite{althaus2014estimating} where the total population
size in each country (Guinea, Sierra Leone and Liberia) was assumed to be
$10^6$ individuals.\\

{\bf 2.} An interesting (and most reliable in our opinion) approach
would be considering "relatively small" number of population $S(0)$
as a variable that needs to be estimated.
We have implemented this approach and the results show that
currently available curve/data is not "long" enough to uniquely determine
$S(0);$ that is, almost the same quality of data fit can be achieved for different
numbers  $S(0)$ (we have tried 10,000, 5000, 100,000 and 200,000) leading
to different numbers of ``stabilized" cumulative infected cases and infection periods.
Taking this factor into account, we do not consider this approach,
however we note that it might be quite possible soon with the availability of more data points.\\

{\bf 3.} In this paper we adopt another approach by neglecting the compartment $S$ completely and leaving just the compartment $I.$
The force of infection $F(S,I) $ in this case is the main factor to be determined.
We take this function in the form
\begin{equation}\label{F}
    F(S,I) = \beta \, \mathcal{D} I
\end{equation}
where $\mathcal{D}$ is the population density of a particular country. In a more general setting,
one would involve functions nonlinear in $I$ (like $F(S,I) = \beta \, \mathcal{D} I^\xi$ with $\xi \le 1$).
However, since the infectious population $I$ is a very small portion of the total population, function $F$ can be assumed linear at least in early stages of  epidemics.
 In this case equation  (\ref{I_A}) can be represented in the form
\begin{equation}\label{I}
     I(t+1) = (1-\mu)^{\tau_1} \beta \, \mathcal{D} I(t-\tau_1) + (1 - \mu - \alpha - \gamma) I(t).
\end{equation}
The major drawback of this model is that $I$ may growth infinitely if the reproduction number is
grater than 1; in this model there is
no variable/parameter (like $S(t)$ in $SIR$) that could force $I$ to decrease.
On the other hand we believe that it can better describe the behavior of an infected population in ``small"
time intervals and provide more accurate reproduction numbers.\\

{\bf Active infectious population.~}
Now we discuss the infectious population and equation (\ref{I}) in more detail. We call ``active infectious
populations" at time $t$ the infected population that are infectious at that time but are not hospitalized yet.
Denote by $I_a (t)$ the number of active infectious
populations at time $t.$ We will rewrite equation (\ref{I}) in terms of $I_a.$

Denote by $\tau_2$ the average time of infectiousness; that is, time from onset ($\tau_1$) to hospitalization. Then, an infected person is assumed to be active infectious during the period $[\tau_1, \tau_1 + \tau_2].$
Since $\tau_2$ is relatively small, we can assume that
none is recovering during that period. This means that the rate of recovery $\gamma$ in (\ref{I}) is no longer needed
for $I_a (t).$

Thus, we transform equation (\ref{I}) by taking into account the time delay $\tau_2.$
Accordingly, the equation for $I_a (t)$ can be represented in the form
\begin{equation}\label{I_a}
     I_a(t+1) =  (1-\mu)^{\tau_1}  \sum_{i = 0}^{\tau_2-1} (1 - \mu)^i (1-\alpha \omega(i)) \, \beta \, \mathcal{D} \, I_a(t-\tau_1 - i) .
\end{equation}
Here $\omega(0)=0$ and $\omega(i), i \ge 1,$ is a gamma cumulative distribution function for onset-to-death that well describes the current Ebola virus in West Africa \cite{periods2014ebola}. We note that in this equation, for each $i\ge 1,$ the fraction $(1-\alpha \omega(i))$ is applied to the remaining infectious $(1 - \mu)^i \, \beta \, \mathcal{D} \, I_a(t-\tau_1 - i);$ that is, the death rate in (\ref{I_a}) is slightly different from (\ref{I})   (indeed, both $\mu$ and $\omega(i)$ are quite small and this leads to $1 - \mu -\alpha \omega(i) \approx (1 - \mu) (1-\alpha \omega(i))$).\\

{\bf Cumulative number of infected cases.~}
The first term $(1-\mu)^{\tau_1} \beta \, \mathcal{D} \, I_a(t-\tau_1)$ in (\ref{I_a})
describes the number of new cases at time $t.$ The cumulative number of
infectious cases at $(t+1)$ will be denoted by $C(t+1).$ It can be calculated as
\begin{equation}\label{C}
     C(t+1) =  (1-\mu)^{\tau_1}  \sum_{s = 1}^{t} \beta \, \mathcal{D} \, I_a(s-\tau_1).
\end{equation}

{\bf Cumulative number of deaths.~}
To calculate the cumulative number of deaths at time $t,$ we consider all infectious cases (hospitalized or not) in the interval
$[t-\tau_1,t-n]$ where $n$ is a sufficiently large number. In particular we assume that death may occur
after the onset. As mentioned above, the distribution of death is described by a gamma
distribution function $\omega$ with its p.d.f - $\omega_p.$ Then, the cumulative number of deaths due to disease can be calculated as
\begin{equation}\label{D}
     D(t+1) =  (1-\mu)^{\tau_1} \sum_{s = 1}^{t}  \sum_{i = 0}^{n} (1 - \mu)^i\, \alpha\, \omega_p(i) \, \beta \, \mathcal{D} \, I_a(s-\tau_1 - i).
\end{equation}

{\bf Basic reproduction number  - $R_0$.~}
We calculate the basic reproduction number by considering stationary infectious $I_a(t) \equiv I^*$ in (\ref{I_a}) as follows:
\begin{equation*}\label{R_0}
    R_0 = \beta \, \mathcal{D} (1-\mu)^{\tau_1}  \sum_{i = 0}^{\tau_2-1} (1 - \mu)^i (1-\alpha \omega(i)),
\end{equation*}
Since the natural death rate $\mu$ is close to zero; that is, $1-\mu \approx 1,$ from (\ref{R_0}) we have
\begin{equation*}\label{R_0_1}
    R_0 \approx \beta \, \mathcal{D} \, [\tau_2 - \alpha \sum_{i = 0}^{\tau_2-1} \omega(i)].
\end{equation*}
Moreover, since $\alpha \sum_{i = 0}^{\tau_2-1} \omega(i) < 1,$ the reproduction number $R_0 \approx \beta \, \mathcal{D} \tau_2.$
This means that the reproduction number very much depends on $\tau_2$ - the time from onset to hospitalization.\\

{\bf The effective reproduction number - $R_k, k \ge 1.$~}
The effective reproduction number $R_k$ on some time interval $[T^k_1,T^k_2]$ can be calculated by the above formula by considering corresponding values of the parameters (mainly, for $\beta$ and $\tau_2$).
The main focus in this case is the study of possible change in the optimal value $\tau_2$ that very much defines the dynamics of the infectious population.

\section{Data fitting: Optimization Problems}
{\bf Main parameters.~} We have formulated the dynamical system (\ref{I_a}),(\ref{C}),(\ref{D}). Given the observed cumulative number of infected cases - $C^0(t)$ and cumulative number of death cases - $D^0(t),$ the parameters of the systems can be estimated by the best fit. Before formulating this problem we discuss the parameters to be estimated.

The density $\mathcal{D}$ and the natural death rate of the population - $\mu$  is available for each country.
We set $\mathcal{D}=41, 80$ and $36$ for Guinea, Sierra Leone and Liberia, respectively. The natural death rate is around 10 deaths for 1000 population per year (1 percent yearly) for all the three countries. Thus in all numerical implementations, the daily rate $\mu$ is set to be $0.01/365 = 0.0000274.$

It is reasonable to have the same average latency period - $\tau_1$ for infected individual to become infectious.
The previous studies (e.g. \cite{periods2014ebola}) suggest that it is between 2-21 days with the mean of 11.4 days.
Our numerical experiments show that $\tau_1 = 8$ is the best for Guinea and $\tau_1 =7$ for both Sierra Leone and Liberia
and $\tau_1 = 6$ for worldwide.

Parameters of the gamma distribution can be taken from \cite{periods2014ebola}. We set
\begin{equation}\label{gamma}
    \omega_p(x) = \frac{b^a}{\Gamma(a)} x^{a-1} e^{-bx}, ~~ a = 10, ~ b = 1.3333
\end{equation}
with mean value 7.5. Note that the choice of values $a$ and $b$ within reasonable intervals, by keeping the mean value the same, has almost no effect on the quality of data fitting. Taking into account this fact, the parameters of the gamma distribution are chosen as in
(\ref{gamma}).
In all the calculations, we set $n=35$ (days) in (\ref{D}) (note that for large $i$ function values $\omega(i)$ are almost zero).

Initial values $I_a(t), ~ t \le 1,$ for the equation are chosen in the form
\begin{equation}\label{I_0}
    I_a(t) = \xi C^0(1), ~ {\rm ~for~all~} t \le 1.
\end{equation}
where $C^0(1)$ is the actual cumulative infectious.
Numerical experiments show that the choice of $\xi$ has very little impact on the quality of data fitting.
The best values for $\xi$ are around
0.6 (Guinea), 0.5 (Sierra Leone), 0.7 (Liberia) and 0.4 (World);
small changes around these values do not effect the results. Taking into account this observation,
we do not consider $\xi$ as a variable and set the above mentioned values
for each country/data.\\

Therefore, the main parameters that define the dynamics of Ebola epidemics in different countries are:
\begin{itemize}
  \item $\alpha$ - the death rate due to disease;
  \item $\beta$ - the coefficient of the force of infection;
  \item $\tau_2$ - the time to hospitalization.
\end{itemize}

\bigskip

{\bf Data fitting.~}
We consider data collected till 11 November 2014 for the cumulative number of infectious (confirmed, probable and suspected) and death individuals; they will be denoted by  $C^0(t)$ and $D^0(t),$  respectively. We will use the root mean square error. Given time interval $[T_1,T_2]$
and data points $C^0(t_i)$ and $D^0(t_i),$ $i \ge 1,$ we  define
\begin{equation}\label{Obj}
    O ([T_1, T_2]) = \sum_{t_i \in [T_1,T_2]} \left[ (C(t_i) - C^0(t_i))^2 + (D(t_i) - D^0(t_i))^2   \right].
\end{equation}

\subsection{Basic reproduction number $R_0$}

\bigskip

To calculate the basic reproduction number, the above model is considered on the whole interval.
The objective function is evaluated on the interval  $[T_1,T_2]$  where in all cases
the last day of data used in the experiments is  $T_2 = $  11-Nov-2014. The starting time  $T_1$
is taken the half of the infectious period in each country; that is,  $T_1 = T_2/2.$
The corresponding data fitting problem is:\\

{\bf Problem ($DF_1$): ~} Given data $C^0(t_i)$ and $D^0(t_i),$ $i \ge 1,$ and time interval $[1,T_2]:$\\

\begin{center}
Minimize $f(\alpha, \beta, \tau_2) = \sqrt{O ([\frac{T_2}{2},T_2])}$;\\ ~~~~ subject to (\ref{I_a})-(\ref{I_0}).\\
\end{center}

\subsection{The reproduction numbers $R_k, k = 1,2,3$ for different time sections}
The reproduction number is mainly determined by $\beta$ and $\tau_2.$
Since in our model parameter $\tau_2$ takes discrete values (days) it would be interesting to study
the change of this parameter over time while keeping $\beta$ the same for a whole period.

We consider three consequent intervals $\Delta_k = [t_k, t_{k+1}]$ $(k = 1,2,3)$ for each country and find optimal values
$\alpha,$ $\beta$ and $\tau_2^k$ $(k = 1,2,3).$
The last time point $t_4$ is $T_2=$ 11-Nov-2014. The values of $t_1, t_2,t_3$ are as follows:
22-March, 23-May and 20-July for Guinea;
27-May, 20-June and 20-August for Sierra Leone;
16-June, 20-July and  07-Sept for Liberia; and
22-March, 23-May and  07-Sept for the total data (World).
Each interval $\Delta_k$ has its own reproduction number $R_k$ that defines the shape of the best fits.

Corresponding data fitting problem is:\\

{\bf Problem ($DF_2$): ~} Given data $C^0(t_i)$ and $D^0(t_i),$ $i \ge 1,$ and time interval $[t_1,t_4]:$
\begin{center}
Minimize $f(\alpha, \beta, \tau_2^1,\tau_2^2,\tau_2^3) = \sqrt{
O ([\frac{t_4-t_1}{2},t_{4}])}$;\\
subject to: (\ref{I_a})-(\ref{I_0}), ~~ where in (\ref{I_a}) ~ $\tau_2 = \tau_2^k, \forall t \in \Delta_k, k=1,2,3.$
\end{center}


\begin{thebibliography}{10}

\bibitem{althaus2014estimating}
C.L. Althaus.
\newblock Estimating the reproduction number of ebola virus (ebov) during the
  2014 outbreak in west africa.
\newblock {\em Plos Currents Outbreaks}, 2014
  (http://currents.plos.org/outbreaks/article/estimating-the-reproduction-numb%
er-of-zaire-ebolavirus-ebov-during-the-2014-outbreak-in-west-africa/).

\bibitem{Bosely}
S.~Boseley.
\newblock Who warns 10,000 new cases of ebola a week are possible.
\newblock {\em The Guardian}, (15 October), 2014,
  http://www.theguardian.com/world/2014/oct/14/who-new-ebola-cases-world-healt%
h-organisation.

\bibitem{browne2014model}
C.~Browne, X.~Huo, P.~Magal, M.~Seydi, O.~Seydi, and G.~Webb.
\newblock A model of the 2014 ebola epidemic in west africa with contact
  tracing.
\newblock {\em arXiv preprint arXiv:1410.3817}, 2014.

\bibitem{chowell2004basic}
G.~Chowell, N.~W Hengartner, C.~Castillo-Chavez, P.W. Fenimore, and J.M. Hyman.
\newblock The basic reproductive number of ebola and the effects of public
  health measures: the cases of congo and uganda.
\newblock {\em Journal of Theoretical Biology}, 229(1):119--126, 2004.

\bibitem{chowell2014transmission}
G.~Chowell and H.~Nishiura.
\newblock Transmission dynamics and control of ebola virus disease (evd): a
  review.
\newblock {\em BMC medicine}, 12(1):196, 2014.

\bibitem{Ganso}
Global and Non-Smooth~Optimization library (GANSO).
\newblock Federation university australia.
\newblock http://www.ganso.com.au.

\bibitem{li2014sir}
M.~Li and X.~Liu.
\newblock An sir epidemic model with time delay and general nonlinear incidence
  rate.
\newblock 2014, 2014 (http://www.hindawi.com/journals/aaa/2014/131257/abs/).

\bibitem{Mam-2005-Chapter}
M.~Mammadov, A.~Rubinov, and J.~Yearwood.
\newblock Dynamical systems described by relational elasticities with
  applications.
\newblock {\em Continuous Optimisation: Current Trends and Modern Applications,
  V.Jeyakumar and A. Rubinov (Eds)}, pages 365--385, 2005.

\bibitem{nishiura2014early}
H.~Nishiura and G.~Chowell.
\newblock Early transmission dynamics of ebola virus disease (evd), west
  africa, march to august 2014.
\newblock {\em Euro Surveill}, 19:36, 2014.

\bibitem{rivers2014modeling}
Caitlin~M Rivers, Eric~T Lofgren, Madhav Marathe, Stephen Eubank, and Bryan~L
  Lewis.
\newblock Modeling the impact of interventions on an epidemic of ebola in
  sierra leone and liberia.
\newblock {\em arXiv preprint arXiv:1409.4607}, 2014.

\bibitem{periods2014ebola}
WHO Ebola~Response Team.
\newblock Ebola virus disease in west africa—the first 9 months of the epidemic
  and forward projections.
\newblock 2014 (http://www.nejm.org/doi/full/10.1056/NEJMoa1411100).

\end{thebibliography}

\end{document}